\documentclass[useAMS,usenatbib]{mn2e}

\usepackage{amsmath,epsfig}

  \let\url\relax

\title{First Detection of Cosmic Structure in the 21-cm Intensity Field}

\author[Pen et al.]{ Ue-Li Pen$^1$, Lister Staveley-Smith$^2$, Jeffrey
  B. Peterson$^3$, Tzu-Ching Chang$^{1,4}$\\ $^{1}$ CITA, University
  of Toronto, 60 St.George St., Toronto, ON, M5S 3H8, Canada \\ $^{2}$
  School of Physics, M013, University of Western Australia, 35
  Stirling Highway, Crawley WA 6009, Australia \\ $^{3}$ Department of
  Physics, Carnegie Mellon University, 500 Forbes Ave, Pittsburgh, PA
  15213, USA \\ $^{4}$ Institute for Astronomy and Astrophysics,
  Academia Sinica, P.O. Box 23-141, Taipei 10617, Taiwan}
\begin{document}

\date{\today}

\maketitle

\begin{abstract}
We present the first statistically significant detection of cosmic
structure using broadly distributed hydrogen radio emission. This is
accomplished using a cross correlation with optical galaxies.
Statistical noise levels of $20 \mu $K are achieved, unprecedented in
this frequency band.  This lends support to the idea that large
volumes of the universe can be rapidly mapped without the need to
resolve individual faint galaxies, enabling precise constraints to
dark energy models.  We discuss strategies for improved intensity
mapping.
\end{abstract}



\section{Introduction}

Many of the outstanding questions in cosmology can be addressed using
maps of structure in the universe. Three dimensional maps can reveal
the dynamics of the expansion, and are especially useful to study the
effects of dark energy.  Previously, the only known way to obtain a 3D
map of the structure in the universe is via galaxy redshift surveys,
that is by isolating millions of individual galaxies, recording
spectra for each and determining each redshift.  Galaxies appear
fainter the further away they are, which makes their measurement
progressively more expensive as the survey extends to higher redshift.

On the other hand, it may be possible to directly map 3D structure at
high redshift by measuring line emission with coarse resolution.  In
this scheme the blended emission of many galaxies is collected
together allowing large scale structure to be studied even though few
galaxies are individually detectable. Here the aggregate emission is
treated as a continuous 3D intensity field so we call this strategy
``Intensity Mapping''.

Intensity mapping has a distinct advantage over redshift survey
techniques because of the use of detection thresholds in redshift
surveys.  Galaxies are typically not entered into a redshift catalog
unless they are detected with high confidence.  Often a $ > 5\sigma$
threshold is set for inclusion.  By using only the $5\sigma$ peaks the
galaxy survey throws away the great majority of the emission. In
contrast, in the intensity mapping scheme one measures correlations
within an intensity field. Here no threshold is needed and all the
emission is used, allowing large volumes to be rapidly surveyed.

If the goal is the study of primordial structure, map resolution finer
than 10 Mpc is not useful.  Features of size smaller than this may
have been imprinted early in the big bang but those early imprints
have been erased by more recent gravitational motions.  At 10 Mpc
resolution a typical 3D map pixel contains many galaxies, making
searches for their aggregate emission promising.

Several recent papers have proposed using Intensity Mapping as a tool
to trace the evolution of dark energy \citep{2007arXiv0709.3672C,2008MNRAS.383..606W}.  Large scale ($>10$
Mpc) structure maps have imprinted in them a small periodic density
variation due to Baryon Acoustic Oscillations (BAO).  These are relics
of the prominent peaks seen in the spatial power spectrum of the
cosmic microwave background.
The baryon oscillation wavelength serves as a comoving "standard
ruler".  Measurement of the angular (and redshift space) wavelengths
of these oscillation as a function of redshift will allow precise
measurement of the kinematics of the expansion.  In most cosmic
evolution models dark energy only becomes dynamically important at
$z<2$, so measurements at such redshifts are needed to distinguish
among these models.

The 21cm hydrogen transition is the dominant spectral line at
frequencies less then 1420 MHz, and is an isolated transition which
allows a direct translation of the frequency of a source into its
redshift (distance).  This means the 21cm transition is well suited
for a three dimensional intensity mapping experiment.

The challenge in such an experiment is to detect cosmic structure in
the 3D 21 cm intensity field beneath the much brighter flux from
continuum sources. Here we show that detection of cosmic structure via
21 cm intensity mapping is indeed possible.  We make this initial
detection by use of a template provided by an optical galaxy redshift
survey.

\section{Data}

The nearby southern sky was mapped in the redshifted 21cm line as part
of the HIPASS survey \citep{2001MNRAS.322..486B}.  This program was
designed under the traditional galaxy redshift survey scheme in which
individual galaxies are detected at high confidence, these positions
are cataloged and the search for cosmic structure uses the catalog.
Here we do not use the HIPASS galaxy catalog (HICAT), but instead use
21 cm spectral intensity data.  The HIPASS survey extends to a
distance of 127 $h^{-1}$ Mpc and covers all declinations south of
+25.5$^\circ$.

\begin{figure}
\centerline{\epsfig{file=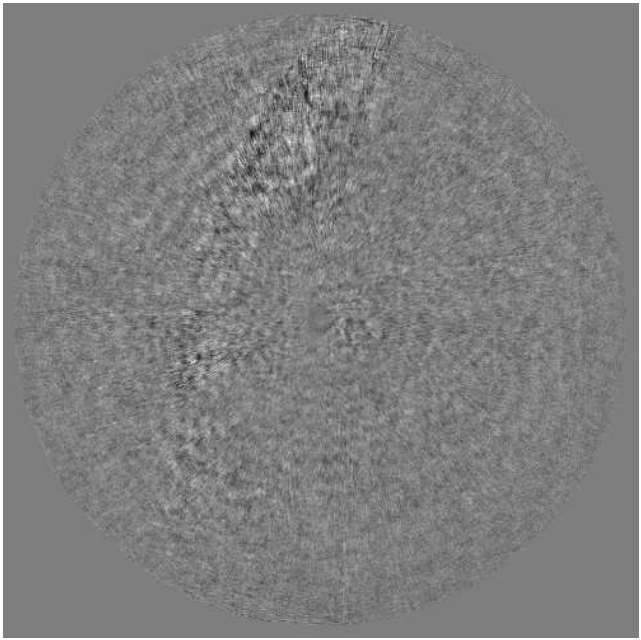, width=\columnwidth}}
\caption{
The HIPASS data cube $R<127 h^{-1}$ Mpc, projected in a cartesian
coordinate system towards the south pole.  
}
\label{fig:hipass}
\end{figure}

\begin{figure}
\centerline{\epsfig{file=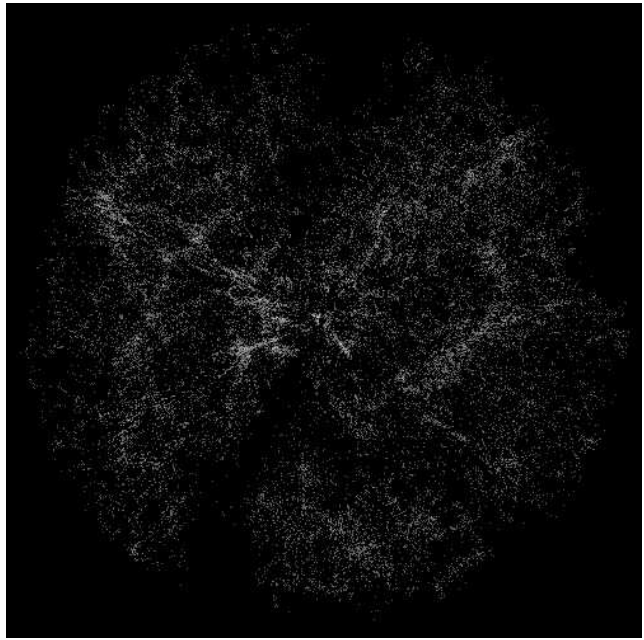, width=\columnwidth}}
\caption{
The 6dFGS catalog for $R<127 h^{-1}$ Mpc, also projected towards the
south pole.  The missing wedges are the
galactic plane. 
}
\label{fig:6df}
\end{figure}

Figure \ref{fig:hipass} shows a cartesian projection of the HIPASS
intensity field.  The region shown is $254 h^{-1}$ Mpc across and
image pixel size corresponds to the observation beam size $0.5 h^{-1}$
Mpc. The large scale structure of the universe as mapped by the
galaxies (see Figure \ref{fig:6df}) is not apparent to the eye in this
21 cm image.  Instead two types of features stand out: ring
structures, likely due to discreteness in the declination scan
strategy, and residual continuum emission from the Milky Way, which
appears as a broad diagonal structure on the left half of the
image. The structure visible in this map is due to cleaning artifacts,
including scan to scan calibration variation and incomplete continuum
source subtraction.
 
Much of the HIPASS volume was also surveyed optically.  The six degree
field galaxy redshift survey (6dFGS;\citet{2004MNRAS.355..747J,2005PASA...22..277J}) is a redshift survey
of the southern sky.  The full 6dFGS contains more than 120,000
redshifts, and was collected from 2001 to 2005.  The majority of
targets were taken from the 2MASS catalog.

6dFGS contains 27417 galaxy redshifts in the region of overlap with
HIPASS.  Their spatial locations are shown in figure \ref{fig:6df}.
Unlike the 21 cm image the typical features of redshift maps are
apparent, including the filamentary structure, and velocity
distortions, which appear radial in this projection. The figure
prominently displays the cosmic web of large scale structure.

\begin{figure}
\centerline{\epsfig{file=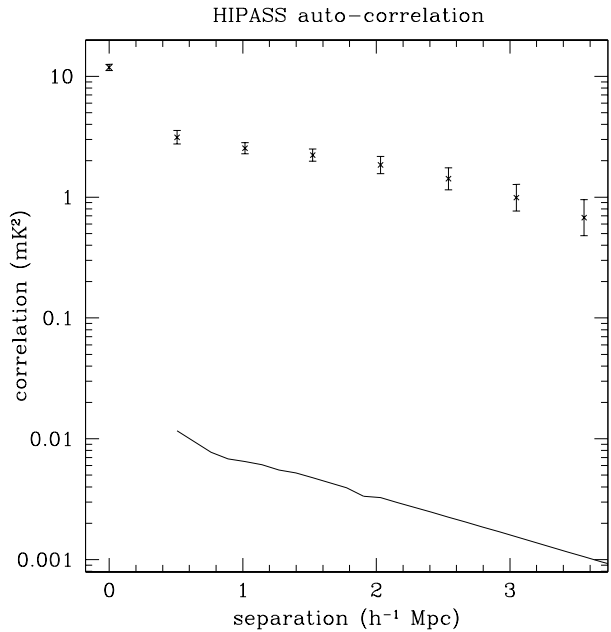, width=\columnwidth}}
\caption{
The auto correlation of the HIPASS data cube, in a shell of thickness
13 $h^{-1}$ Mpc, centered at 95 $h^{-1}$ Mpc distance.  The solid line
is the model prediction (see text).  The signal is due to residual
foreground subtraction.  The vertical axis is a variance, so the
residual foreground fluctuations of $\sim m$K is about a factor of ten
larger than the expected large scale structure signal.
}
\label{fig:auto}
\end{figure}

Just as the eye is drawn to the continuum residuals in figure
\ref{fig:hipass}, the autocorrelation of the HIPASS data is dominated
by spurious correlations due to these structures.  Sampled with the
0.5 $h^{-3}$ Mpc$^3$ pixels we used, the variance is about (3.6
mK)$^2$.  If these variations were uncorrelated between pixels, this
would average down when larger structures were examined.  For example,
when rebinned to cells of the nonlinear length scale, i.e. cells of
$\sim$10 $h^{-1}$ Mpc radius, uncorrelated noise would go down by
$\sqrt{ \sim 8000}$, resulting in a pixel noise around 40 $\mu$K.
This would in principle yield a signal-to-noise comparable to the
sample variance, leading do a sample variance limited power spectrum.
Over the HIPASS survey volume, there are 1000 such cells, which could
lead to a 30 $\sigma$ detection of the large scale structure signal in
the autocorrelation function.  In practice, however, the noise does
not reduce as one averages on larger scales.  Figure \ref{fig:auto}
shows the measured autocorrelation in a redshift shell of thickness
13 $h^{-1}$ Mpc, centered at 95 $h^{-1}$ Mpc distance.  The crosses with
error bars are the measured correlation, which is much larger than the
expected signal (solid line, described below).  The excess variance is
presumably due to the residual from incomplete foreground subtraction.

The HIPASS observing strategy was designed to detect galaxies, rather
than extended 21 cm structures.  To bring out the cosmic structure
despite the continuum artifacts we computed the cross correlation of
the 21 cm intensity field with the positions of the optically detected
galaxies. In the limit of a large survey volume, in such a cross
correlation, the contribution from radio continuum residuals should
average to zero, as long as the optical and radio surveys do not share
common artifacts. If there are no noise correlations between the two surveys,
averaging over the $N =$ 27417 6dFGS galaxies, one
expects the statistical noise to reduce by $\sqrt{N}$ to 22 $\mu$K.

\section{Cross-Correlation}

We computed a radio-optical cross correlation, starting by gridding
both the 6dFGS galaxies and the HIPASS data on a regular grid, 500
grid cells wide, with grid spacing of 0.5 $h^{-1}$ Mpc.  Our goal is
to detect cosmic structure in the portion of the survey where
individual galaxies are difficult to detect with confidence. We only
used data with distances greater than 63 $h^{-1}$ Mpc, i.e. distances
at least half way to the survey edge.  This includes the majority
(7/8th) of the survey volume but contains only 7\% of the 4315
detected HICAT galaxies. 

The HIPASS data we used had already been filtered both spatially and
spectrally.  The data were recorded while sweeping the telescope in
declination, and the median of neighboring strips in declination was
subtracted from each beam-width region.  This spatial filter
suppresses structures in the declination direction bigger than the
beam, which is about 0.5 $h^{-1}$ Mpc at a typical distance of 100
$h^{-1}$ Mpc.  In order to measure a largely unsuppressed correlation,
we only included pairs of HIPASS and 6dFGS cells with a transverse
separation of less than 0.7 $h^{-1}$ Mpc.  The frequency domain was
also filtered to remove continuum point sources. This is accomplished
by removing all flux from each pixel that fits a low order
polynomial. Because of the slow variation with frequency of the
spectra of the continuum sources they are strongly attenuated, leaving
behind the 21 cm line emission.

\begin{figure}
\centerline{\epsfig{file=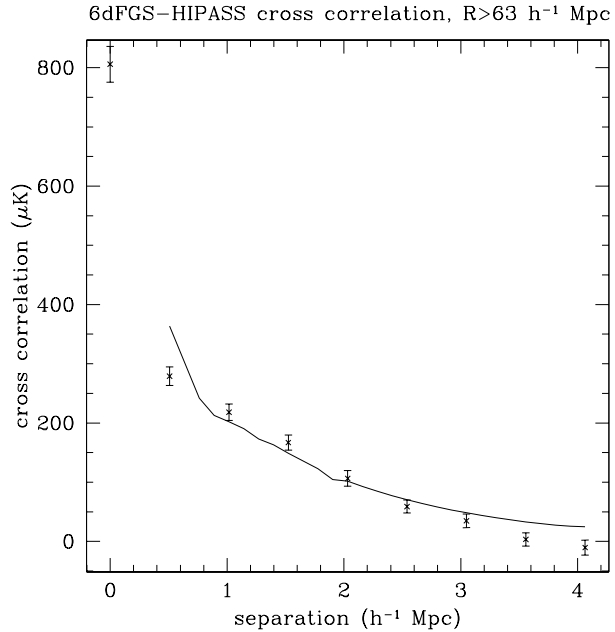, width=\columnwidth}}
\caption{ The correlation function of hydrogen 21 cm emission when
  stacked around optically selected galaxies from the 6dFGS.  The
  solid line is a standard clustering model.  The point at zero
  separation is the mean flux from 6dF galaxies, while points at
  larger separations measure the associated surrounding large scale
  structure.  HI emission from the 6dF galaxies could spill up to 1 $h^{-1}$
  Mpc into neighboring bins at the $\sim 20 \mu$K level.  }
\label{fig:corr}
\end{figure}

Figure \ref{fig:corr} shows the measured cross correlation.  We see a
significant detection at separations up to 3 $h^{-1}$ Mpc.  At larger
separation the correlation signal is strongly suppressed by the
spatial filtering.  The signal is a measure of the collective HI
emission of volumes adjacent to 6dFGS galaxies.  The strong cross
correlation at zero separation is the HI emission from the 6dFGS
galaxies themselves.  The solid line is an estimate of the expected
correlation, using the 2dFGS model \citep{2003MNRAS.344..847M}.
This model adopts a power law form for the cross correlation function
$\xi=35 (r/r_0)^{-1.8} \mu$K with a standard correlation length $r_o
=$ 5.5 $h^{-1}$ Mpc \citep{1993ppc..book.....P}.  6dFGS galaxies are
preferentially early type, and typically have a longer correlation
length than our choice for $r_0$.  HIPASS sources are preferentially
gas rich late type, which typically have a shorter correlation
length. We selected a mean value.  A bias of $b=1$ was assumed.  The
average 21 cm sky brightness was taken as 35 $\mu$K
\citep{2007arXiv0709.3672C}, using $\Omega_{\rm HI}=0.0004$
\citep{2005MNRAS.359L..30Z}.  A radial velocity dispersion was also
assumed, which is convolved with the correlation function. This
dispersion value was the only free parameter, chosen to fit the
data. The best fit pairwise dispersion is 200 km/sec. The power law
correlation model does not work at zero separation, so that data point
was excluded from the fit.

The noise level of the cross correlation is less than 20 $\mu$K. This
uncertainty was determined by subdividing the data set and comparing
the results of the subsets.  A direct measurement of the BAO signal
was made by \citet{2005ApJ...633..560E}, at a correlation amplitude
wiggle of $\sim 0.01$ and a lag of $109 h^{-1}$ Mpc.  In the units of
figure \ref{fig:corr}, this corresponds to an amplitude of $\sim 0.3
\mu$K, so a factor of 30 larger in separation and a factor of 100
smaller in the current achieved noise amplitude.  The maximum radius
of the HIPASS survey is too small to search for features with such a
large wavelength.

\section{Future Prospects}

A program to precisely measure baryon
oscillation--and thereby to study dark energy
\citep{detf,2007ApJ...665...14S}--requires a volume $10^4$ that of
the HIPASS data set.  This would require a dedicated intensity mapping
survey program, and the success reported here may encourage 21 cm
observers to rapidly proceed with such a program.  There will be no
optical survey of such a large volume available for cross correlation
for many years to come.  However, while waiting for these surveys it
still may be possible to detect and study cosmic structure in the 21
cm intensity field by autocorrelation. This will require much more
complete continuum removal than has been accomplished so far in the
HIPASS data set.

Several programs\footnote{Giant
  Metrewave Radio Telescope (GMRT;
  {\url{http://www.ncra.tifr.res.in}}), Low Frequency Array (LOFAR;
  \url{http://www.lofar.org}), Murchison Widefield Array (MWA;
      {\url{http://web.haystack.mit.edu/arrays/MWA}}), Primeval
      Structure Telescope (PAST;
      \url{http://web.phys.cmu.edu/$\sim$past/}), and Square Kilometre
      Array (SKA; {\url{http://www.skatelescope.org}}).} are already underway to measure the collective 21cm
emission during the epoch of reionization, $z>6$, when the
neutral fraction of the universe was closer to unity. At the low
frequency of these observations the continuum sources are very bright,
so under these programs effective tools for continuum removal are
being developed. 

In principle, the Baryon Acoustic Oscillation signal could be detected
at redshifts near that of the reionization.  Assuming the universe was
matter dominated at such high redshifts, the position of the BAO peaks
is directly tied to the position of the peaks at the CMB power spectrum. 
The BAO peak positions should be redshift independent between $z\sim 1100$ and $z\sim 6$.  Any
deviation from this expected peak position would amount to
detection of dynamical influences of dark energy at very early
times. This is unexpected, and so would be intriguing if found.

A low ($z<2$) redshift 21 cm Intensity Mapping program would share
many features with CMB surveys, which do detect baryon oscillation via
autocorrelation (see \citet{2007RPPh...70..899T} for a recent review,
and references therein).  On existing single-dish 21 cm telescopes,
the observing strategy would be to scan in azimuth at constant
elevation, but to do this for each field at a several local hour
angles, thereby creating a cross-linked data set. Such observing
techniques and their analysis pipelines have developed over the forty
year history of CMB observations, and are very effective
\citep{2007ApJS..170..377S,2008arXiv0801.1491R,2006ApJ...647..823J}. Such
a program might succeed in detecting the cosmic web structure but the sky coverage
needed for precise BAO studies may be difficult to obtain using
existing single dish telescopes.

One approach to rapidly map large areas of sky is to use
interferometers, which offer excellent instrumental stability and
mapping speed.  Since one is searching for large angular scale
features, existing interferometers such as the Very Large Array, are
not appropriate since they are insensitive to large angular scales.
Instead, a dedicated compact array, in the spirit of successful CMB
experiments like DASI \citep{2002Natur.420..772K} and CBI
\citep{2002PASP..114...83P}, could achieve an all sky map of most of
the visible universe \citep{2007arXiv0709.3672C,2008MNRAS.383..606W}.

\section{Conclusions}

We report the first detection of cosmic structure via extended 21 cm
hydrogen emission.  Optically measured galaxy positions in three
dimensions were used as a finding chart to locate the density peaks of
cosmic structure, and the 21 cm excess brightness is detected at up to
3 $h^{-1}$ Mpc distance from these galaxies.  Due to the spatial
filtering in the HIPASS maps, most of the correlation signal comes
from correlations along the line of sight, rather than transverse 
structure.  The detection is a cross correlation, and represents a
first step towards mapping the cosmic structure using 21cm emission.

Proposed programs to measure baryon oscillations via 21 cm intensity
mapping will need to measure autocorrelations of 21 cm flux at
separations larger than 100 $h^{-1}$ Mpc. While the HIPASS data set we used covers
a volume too small to test such plans directly, the low noise level of
the correlation we measure is encouraging.

\bibliography{paper}

\begin{thebibliography}{}

\bibitem[\protect\citeauthoryear{{Barnes}, {Staveley-Smith}, {de Blok},
  {Oosterloo}, {Stewart}, {Wright} \& {et al.}}{{Barnes}
  et~al.}{2001}]{2001MNRAS.322..486B}
{Barnes} D.~G.,  {Staveley-Smith} L.,  {de Blok} W.~J.~G.,  {Oosterloo} T.,
  {Stewart} I.~M.,  {Wright} A.~E.,    {et al.} 2001, \mnras, 322, 486

\bibitem[\protect\citeauthoryear{{Chang}, {Pen}, {Peterson} \&
  {McDonald}}{{Chang} et~al.}{2007}]{2007arXiv0709.3672C}
{Chang} T.-C.,  {Pen} U.-L.,  {Peterson} J.~B.,    {McDonald} P.,  2007, ArXiv
  e-prints, 0709.3672

\bibitem[\protect\citeauthoryear{{Eisenstein}, {Zehavi}, {Hogg}, {Scoccimarro},
  {Blanton}, {Nichol}, {Scranton} \& {et al.}}{{Eisenstein}
  et~al.}{2005}]{2005ApJ...633..560E}
{Eisenstein} D.~J.,  {Zehavi} I.,  {Hogg} D.~W.,  {Scoccimarro} R.,  {Blanton}
  M.~R.,  {Nichol} R.~C.,  {Scranton} R.,    {et al.} 2005, \apj, 633, 560

\bibitem[\protect\citeauthoryear{{Jones}, {Saunders}, {Colless}, {Read},
  {Parker}, {Watson}, {Campbell} \& {et al.}}{{Jones}
  et~al.}{2004}]{2004MNRAS.355..747J}
{Jones} D.~H.,  {Saunders} W.,  {Colless} M.,  {Read} M.~A.,  {Parker} Q.~A.,
  {Watson} F.~G.,  {Campbell} L.~A.,    {et al.} 2004, \mnras, 355, 747

\bibitem[\protect\citeauthoryear{{Jones}, {Saunders}, {Read} \&
  {Colless}}{{Jones} et~al.}{2005}]{2005PASA...22..277J}
{Jones} D.~H.,  {Saunders} W.,  {Read} M.,    {Colless} M.,  2005, Publications
  of the Astronomical Society of Australia, 22, 277

\bibitem[\protect\citeauthoryear{{Jones}, {Ade}, {Bock}, {Bond}, {Borrill},
  {Boscaleri}, {Cabella}, {Contaldi} \& {et al.}}{{Jones}
  et~al.}{2006}]{2006ApJ...647..823J}
{Jones} W.~C.,  {Ade} P.~A.~R.,  {Bock} J.~J.,  {Bond} J.~R.,  {Borrill} J.,
  {Boscaleri} A.,  {Cabella} P.,  {Contaldi} C.~R.,    {et al.} 2006, \apj,
  647, 823

\bibitem[\protect\citeauthoryear{{Kolb} \& {et al.}}{{Kolb} \& {et
  al.}}{2006}]{detf}
{Kolb} E.,  {et al.} 2006,
  http://www.nsf.gov/mps/ast/aaac/dark\_energy\_task\_force/report/
  detf\_final\_report.pdf

\bibitem[\protect\citeauthoryear{{Kovac}, {Leitch}, {Pryke}, {Carlstrom},
  {Halverson} \& {Holzapfel}}{{Kovac} et~al.}{2002}]{2002Natur.420..772K}
{Kovac} J.~M.,  {Leitch} E.~M.,  {Pryke} C.,  {Carlstrom} J.~E.,  {Halverson}
  N.~W.,    {Holzapfel} W.~L.,  2002, \nat, 420, 772

\bibitem[\protect\citeauthoryear{{Madgwick}, {Hawkins}, {Lahav}, {Maddox},
  {Norberg}, {Peacock}, {Baldry}, {Baugh} \& {et al.}}{{Madgwick}
  et~al.}{2003}]{2003MNRAS.344..847M}
{Madgwick} D.~S.,  {Hawkins} E.,  {Lahav} O.,  {Maddox} S.,  {Norberg} P.,
  {Peacock} J.~A.,  {Baldry} I.~K.,  {Baugh} C.~M.,    {et al.} 2003, \mnras,
  344, 847

\bibitem[\protect\citeauthoryear{{Padin}, {Shepherd}, {Cartwright}, {Keeney},
  {Mason}, {Pearson} \& {et al.}}{{Padin} et~al.}{2002}]{2002PASP..114...83P}
{Padin} S.,  {Shepherd} M.~C.,  {Cartwright} J.~K.,  {Keeney} R.~G.,  {Mason}
  B.~S.,  {Pearson} T.~J.,    {et al.} 2002, \pasp, 114, 83

\bibitem[\protect\citeauthoryear{{Peebles}}{{Peebles}}{1993}]{1993ppc..book...%
..P}
{Peebles} P.~J.~E.,  1993, {Principles of physical cosmology}.
Princeton Series in Physics, Princeton, NJ: Princeton University Press, |c1993

\bibitem[\protect\citeauthoryear{{Reichardt}, {Ade}, {Bock}, {Bond}, {Brevik},
  {Contaldi} \& {et al.}}{{Reichardt} et~al.}{2008}]{2008arXiv0801.1491R}
{Reichardt} C.~L.,  {Ade} P.~A.~R.,  {Bock} J.~J.,  {Bond} J.~R.,  {Brevik}
  J.~A.,  {Contaldi} C.~R.,    {et al.} 2008, ArXiv e-prints, 801

\bibitem[\protect\citeauthoryear{{Seo} \& {Eisenstein}}{{Seo} \&
  {Eisenstein}}{2007}]{2007ApJ...665...14S}
{Seo} H.-J.,  {Eisenstein} D.~J.,  2007, \apj, 665, 14

\bibitem[\protect\citeauthoryear{{Spergel}, {Bean}, {Dor{\'e}}, {Nolta},
  {Bennett}, {Dunkley}, {Hinshaw}, {Jarosik} \& {et al.}}{{Spergel}
  et~al.}{2007}]{2007ApJS..170..377S}
{Spergel} D.~N.,  {Bean} R.,  {Dor{\'e}} O.,  {Nolta} M.~R.,  {Bennett} C.~L.,
  {Dunkley} J.,  {Hinshaw} G.,  {Jarosik} N.,    {et al.} 2007, \apjs, 170, 377

\bibitem[\protect\citeauthoryear{{Tristram} \& {Ganga}}{{Tristram} \&
  {Ganga}}{2007}]{2007RPPh...70..899T}
{Tristram} M.,  {Ganga} K.,  2007, Reports of Progress in Physics, 70, 899

\bibitem[\protect\citeauthoryear{{Wyithe} \& {Loeb}}{{Wyithe} \&
  {Loeb}}{2008}]{2008MNRAS.383..606W}
{Wyithe} J.~S.~B.,  {Loeb} A.,  2008, \mnras, 383, 606

\bibitem[\protect\citeauthoryear{{Zwaan}, {Meyer}, {Staveley-Smith} \&
  {Webster}}{{Zwaan} et~al.}{2005}]{2005MNRAS.359L..30Z}
{Zwaan} M.~A.,  {Meyer} M.~J.,  {Staveley-Smith} L.,    {Webster} R.~L.,  2005,
  \mnras, 359, L30

\end{thebibliography}
\bibliographystyle{mn2e}

\section*{Acknowledgments}
 
We acknowledge financial support by NSERC and NSF.

\end{document}